\documentclass[letterpaper,english,american,numberedappendix]{emulateapj}
\usepackage[T1]{fontenc}
\usepackage{amssymb}
\usepackage{graphicx}

\makeatletter

\usepackage{amsmath}
\slugcomment{}
\shorttitle{Stability of Burning on Accreting Neutron Stars}
\shortauthors{Keek, Cyburt, \& Heger}

\makeatother

\usepackage{babel}
\begin{document}

\title{Reaction Rate and Composition Dependence of the Stability of Thermonuclear
Burning on Accreting Neutron Stars}

\author{L.~Keek}

\affil{National Superconducting Cyclotron Laboratory, Department of Physics
\& Astronomy, and Joint Institute for Nuclear Astrophysics, Michigan
State University, East Lansing, MI 48824, USA}

\affil{Center for Relativistic Astrophysics, School of Physics, Georgia
Institute of Technology, Atlanta, GA 30332, USA}

\email{l.keek@gatech.edu}

\author{R.\,H.~Cyburt}

\affil{National Superconducting Cyclotron Laboratory and Joint Institute
for Nuclear Astrophysics, Michigan State University, East Lansing,
MI 48824, USA}

\author{A.~Heger}

\affil{Monash Center for Astrophysics, School of Mathematical Sciences,
Monash University, Victoria, 3800, Australia}

\affil{School of Physics and Astronomy, University of Minnesota, Minneapolis,
MN, 55455, USA}
\begin{abstract}
The stability of thermonuclear burning of hydrogen and helium accreted
onto neutron stars is strongly dependent on the mass accretion rate.
The burning behavior is observed to change from Type I X-ray bursts
to stable burning, with oscillatory burning occurring at the transition.
Simulations predict the transition at a ten times higher mass accretion
rate than observed. Using numerical models we investigate how the
transition depends on the hydrogen, helium, and CNO mass fractions
of the accreted material, as well as on the nuclear reaction rates
of $3\alpha$ and the hot-CNO breakout reactions $^{15}\mathrm{O}\left(\alpha,\gamma\right)\mathrm{^{19}Ne}$
and $^{18}\mathrm{Ne}\left(\alpha,p\right)\mathrm{^{21}Na}$. For
a lower hydrogen content the transition is at higher accretion rates.
Furthermore, most experimentally allowed reaction rate variations
change the transition accretion rate by at most $10\,\%$. A factor
ten decrease of the $^{15}\mathrm{O}\left(\alpha,\gamma\right)\mathrm{^{19}Ne}$
rate, however, produces an increase of the transition accretion rate
of $35\,\%$. None of our models reproduce the transition at the observed
rate, and depending on the true $^{15}\mathrm{O}\left(\alpha,\gamma\right)\mathrm{^{19}Ne}$
reaction rate, the actual discrepancy may be substantially larger.
We find that the width of the interval of accretion rates with marginally
stable burning depends strongly on both composition and reaction rates.
Furthermore, close to the stability transition, our models predict
that X-ray bursts have extended tails where freshly accreted fuel
prolongs nuclear burning.
\end{abstract}


\section{Introduction}

\label{sec:Introduction} The thin envelope of neutron stars in low-mass
X-ray binaries (LMXBs) is continuously replenished by Roche-lobe overflow
of the companion star. The hydrogen- and helium-rich material is quickly
compressed, and after mere hours the density and temperature required
for thermonuclear burning can be reached \citep{Woosley1976,Maraschi1977,Joss1977,Lamb1978}.
If a thermonuclear runaway ensues, the unstable burning engulfs the
entire atmosphere, consuming most hydrogen and helium within seconds.
This powers the frequently observed Type~I X-ray bursts (\citealt{Grindlay1976,1976Belian};
see also \citealt{Cornelisse2003,Galloway2008catalog}; for reviews
see \citealt{Lewin1993,Strohmayer2006}). 

For LMXBs accretion rates, $\dot{M}$, are inferred of up to the Eddington
limit of approximately $\dot{M}_{\mathrm{Edd}}\sim10^{-8}\, M_{\odot}\mathrm{year^{-1}}$
(see Section~\ref{sec:methods}). At high rates close to this limit,
the high heating rate from compression and nuclear burning as well
as the fast inflow of new fuel allow for steady-state burning of hydrogen
and helium (e.g., \citealt{Fujimoto1981,Bildsten1998}). A lower burst
rate and ultimately an absence of bursts is observed at increasingly
large $\dot{M}$, roughly between $0.1\dot{M}_{\mathrm{Edd}}$ and
$0.3\dot{M}_{\mathrm{Edd}}$ \citep{Paradijs1988,Cornelisse2003}.
When the burst rate is reduced, the presence of steady-state burning
becomes evident from an increase of the $\alpha$ parameter, i.e.,
the ratio of the persistent X-ray fluence between subsequent bursts
and the burst fluence: there is an increase in the fraction of fuel
that burns in a stable manner \citep{Paradijs1988}. Understanding
the burning regimes at different $\dot{M}$ allows us to accurately
predict the composition of the burning ashes that form the neutron
star crust, which has observable consequences for, e.g., the cooling
of X-ray transients \citep[e.g.,][]{Schatz2013}.

Whereas observations place the transition of stability around $0.1\dot{M}_{\mathrm{Edd}}$
to $0.3\dot{M}_{\mathrm{Edd}}$ , models predict it to occur at a
mass accretion rate, $\dot{M}_{\mathrm{st}}$, close to $\dot{M}_{\mathrm{Edd}}$
\citep{Fujimoto1981}. The observed $\dot{M}$ is determined from
the persistent X-ray flux. As material from the companion star falls
to the neutron star, most of the rotational and gravitational energy
is dissipated at the inner region of the accretion disk and at a boundary
layer close to the neutron star surface. This causes these regions
to thermally emit soft X-rays, and Compton scattering in a corona
is thought to produce X-rays in the classical band \citep[e.g.,][]{Done2007review}.
The broad-band X-ray flux is, therefore, used to infer $\dot{M}$.
There is some uncertainty in the efficiency of converting the liberated
gravitational potential energy to X-rays, as well as obscuration of
the X-ray emitting regions by the disk. These uncertainties, however,
are generally believed to be at most several tens of percents, whereas
the discrepancy is close to an order of magnitude \citep[see also the discussion in][]{kee06}.
This discrepancy is one of the main challenges for neutron star envelope
models.

At the transition, nuclear burning is marginally stable and produces
oscillations in the light curve \citep{Heger2005}. This has been
identified with mHz quasi-periodic oscillations (mHz QPOs) observed
from hydrogen-accreting neutron stars, which typically occur at accretion
rates close to $0.1\,\dot{M}_{\mathrm{Edd}}$ \citep{Revnivtsev2001,Altamirano2008,Linares2011}.

In the neutron star envelope hydrogen burns through the hot-CNO cycle
\citep[e.g.,][]{Wallace1981}, and helium burns through the $3\alpha$
process. At temperatures above $T\gtrsim5\times10^{8}\,\mathrm{K}$,
breakout from the CNO cycle occurs through the $^{15}\mathrm{O}\left(\alpha,\gamma\right)\mathrm{^{19}Ne}$
reaction, and for $T\gtrsim6\times10^{8}\,\mathrm{K}$ through $^{18}\mathrm{Ne}\left(\alpha,p\right)\mathrm{^{21}Na}$.
This is followed by long chains of $(\alpha,p)$ and $(p,\gamma)$
reactions (the \textsl{$\alpha$p}-process; \citealt{vanWormer1994})
as well as $(p,\gamma)$ reactions and $\beta$-decays (the \textsl{rp}-process).
Isotopes are produced with mass numbers as high as $108$ (\citealt{Schatz2001};
for further discussion about the end point see \citealt{Koike2004,Elomaa2009}).
Detailed numerical studies implement these processes in large nuclear
networks \citep[e.g.,][]{Woosley2004,Fisker2008,Jose2010}. The importance
of key nuclear reactions, such as $^{15}\mathrm{O}\left(\alpha,\gamma\right)\mathrm{^{19}Ne}$,
has been demonstrated for the stability of nuclear burning \citep{Fisker2006,Cooper2006,Fisker2007,Parikh2008,Davids2011,Keek2012}.
Especially for the two breakout reactions the rates are poorly constrained
by nuclear experiment \citep{Davids2011,Mohr2013}, and experimental
work to improve this is ongoing \citep[e.g.,][]{Tan2007,Tan2009,Salter2012,He2013}.

In this paper we investigate the dependence of $\dot{M}_{\mathrm{st}}$
on the reaction rates of the $3\alpha$-process and the CNO-cycle
breakout reactions $^{15}\mathrm{O}\left(\alpha,\gamma\right)\mathrm{^{19}Ne}$
and $^{18}\mathrm{Ne}\left(\alpha,p\right)\mathrm{^{21}Na}$, as well
as on the composition of the accreted material.

\section{Numerical Methods}

\label{sec:methods}

The multi-zone simulations of the neutron star envelope presented
in this paper are created with the one-dimensional implicit hydrodynamic
code KEPLER \citep{Weaver1978}. Nuclear burning is implemented using
a large adaptive network \citep{Rauscher2003}. Two sets of simulations
are made with different versions of KEPLER, both of which have been
used in previous similar studies \citep{Woosley2004,Heger2005,Heger2007,Keek2012}.
Here we describe the main features of these simulations, and we refer
to previous publications for a complete description of the code. 

We model the envelope on top of a neutron star with a $1.4\, M_{\odot}$
gravitational mass and a radius of $10\,\mathrm{km}$. No general
relativistic corrections are applied, but the Newtonian gravity in
our model is the same as the general relativistic gravitational acceleration
in the rest frame of the surface of a star of equal gravitational
mass with a radius of $11.2\,\mathrm{km}$. The corresponding gravitational
redshift of $1+z\simeq1.26$ is \emph{not} applied to the presented
results and light curves \citep[see also][]{Keek2011}. This allows
for easier translation to other choices of neutron star properties
with the same local gravity. Note, however, that the corrected $\dot{M}$
for an observer at infinity differs from our model value by less than
a percent \citep{Keek2011}.

The inner part of our model consists of a $2\times10^{25}\,\mathrm{g}$
$^{56}$Fe substrate, which acts as an inert thermal buffer. At the
bottom boundary of that layer we set a constant luminosity into the
envelope originating from heating by electron capture and pycnonuclear
reactions in the crust \citep{Haensel1990,Haensel2003,Gupta2007}.
For each set of simulations we assume a fixed amount of heat per accreted
nucleon, $Q_{\mathrm{b}}$, enters the envelope. As such the luminosity
at the inner boundary is proportional to the mass accretion rate:
\foreignlanguage{english}{$L_{\mathrm{crust}}=Q_{\mathrm{b}}\dot{M}$}.

On top of the substrate, we accrete hydrogen- and helium-rich material.
We express $\dot{M}$ as a fraction of the Eddington limited mass
accretion rate for material of solar composition, $\dot{M}_{\mathrm{Edd}}\equiv1.75\times10^{-8}\, M_{\odot}\mathrm{yr^{-1}}$.
For easier comparison of values between different models, we use this
rate even when the composition deviates from solar.

\subsection{Models with nuclear reaction rate variation\label{sub:Models-with-rate}}

To study the effect of key nuclear reaction rates on $\dot{M}_{\mathrm{st}}$,
we create models where individual rates are varied within the experimental
uncertainties. We use the thermonuclear reaction rate compilation
REACLIB $2.0$ \citep{Cyburt2010}. In particular, $^{15}\mathrm{O}\left(\alpha,\gamma\right)\mathrm{^{19}Ne}$
is taken from \citeauthor{Davids2011} (\citeyear{Davids2011}; DC11),
$^{18}\mathrm{Ne}\left(\alpha,p\right)\mathrm{^{21}Na}$ from \citeauthor{Matic2009}
(\citeyear{Matic2009}; MV09), and $3\alpha$ from \citeauthor{Caughlan1988}
(\citeyear{Caughlan1988}; CF88). Note that revised formulations of
the $3\alpha$ rate exist, but within the temperature range relevant
for our simulations the rate is dominated by resonant capture, and
the difference is at most $4\%$ \citep[e.g.,][]{Fynbo2005}. 
\begin{figure}
\includegraphics{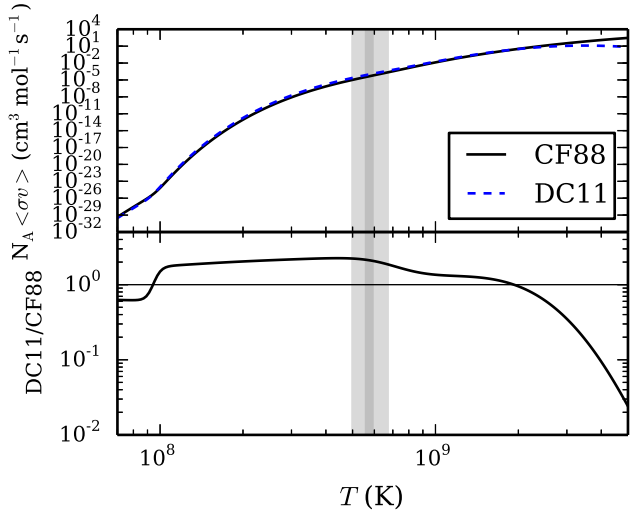}

\caption{\label{fig:rate_o15ag}The temperature dependent part of the $^{15}\mathrm{O}\left(\alpha,\gamma\right)\mathrm{^{19}Ne}$
reaction rate, $N_{\mathrm{A}}\left\langle \sigma v\right\rangle $,
as a function of temperature, $T$, (top), and the ratio of two rates
(bottom) from \citeauthor{Caughlan1988} (\citeyear{Caughlan1988};
CF88) and \citeauthor{Davids2011} (\citeyear{Davids2011}; DC11).
The gray bands indicate the temperature ranges from two models of
marginally stable burning (Figure~\ref{fig:osc_flow_26}), and the
horizontal line helps guide the eye for a ratio of $1$.}
\end{figure}
\begin{figure}
\includegraphics{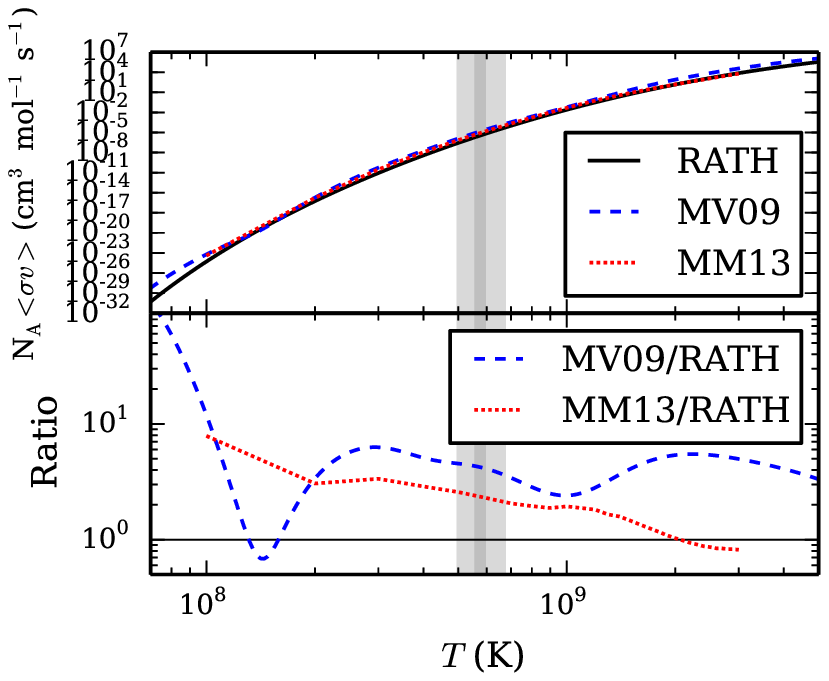}

\caption{\label{fig:rate_ne18ap}Same as Figure~\ref{fig:rate_o15ag} for
the $^{18}\mathrm{Ne}\left(\alpha,p\right)\mathrm{^{21}Na}$ reaction
rate from \citeauthor{Rauscher2000} (\citeyear{Rauscher2000}; RATH),
\citeauthor{Matic2009} (\citeyear{Matic2009}; MV09), and \citeauthor{Mohr2013}
(\citeyear{Mohr2013}; MM13).}
\end{figure}
 
\begin{figure*}
\begin{centering}
\includegraphics{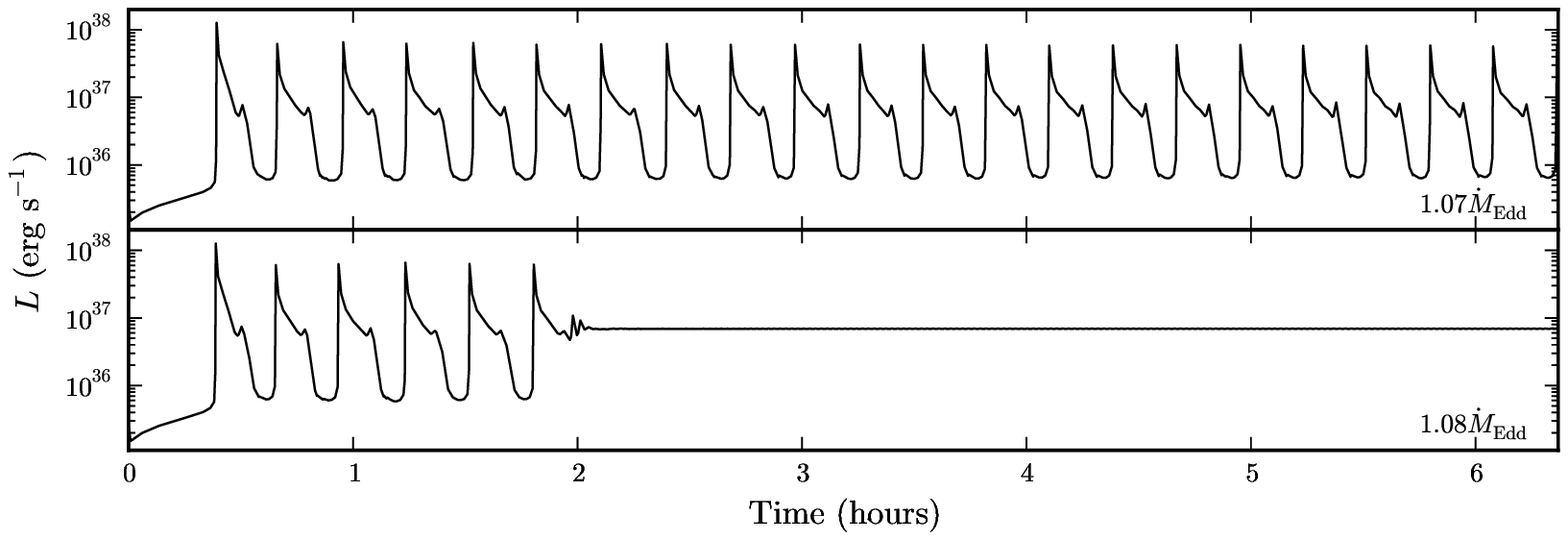}
\par\end{centering}

\caption{\label{fig:standard_lcv}Light curves for models closest to the transition
to stability, with standard reaction rates and indicated mass accretion
rates. The simulation in the lower panel initially exhibits bursts,
but quickly transitions to stable burning. The bursts exhibit extended
tails.}
\end{figure*}
Figures \ref{fig:rate_o15ag} and \ref{fig:rate_ne18ap} illustrate
the $^{15}\mathrm{O}\left(\alpha,\gamma\right)\mathrm{^{19}Ne}$ and
$^{18}\mathrm{Ne}\left(\alpha,p\right)\mathrm{^{21}Na}$ rates, and
compare them to the rates used in our second set of models (Section
\ref{sub:Models-with-composition}), as well as to the recent study
by \citet{Mohr2013} for the latter rate.  We both increase and decrease
the rates of the two CNO breakout reactions by a factor $10$ \citep[e.g.,][]{Davids2011,Mohr2013},
and the $3\alpha$ rate by $20\,\%$ \citep[e.g.,][]{Austin2005},
which we regard as the respective ranges of values allowed by nuclear
experiment.

The outer zone of the models has a mass of $10^{16}\,\mathrm{g}$,
and we use $Q_{\mathrm{b}}=0.1\,\mathrm{MeV\, nucleon^{-1}}$. Accretion
is simulated by increasing the mass and updating the composition of
the zones that form the outer $2\times10^{20}\,\mathrm{g}$ of the
model, and compressional heating is taken into account \citep{Keek2011}.
These zones are close to the surface, well above the depth where hydrogen
and helium burning takes place. The accreted material is of solar
composition with mass fractions of $X=0.71$ ($^{1}$H), $Y=0.27$
($^{4}$He), and $Z=0.02$ ($^{14}$N). The latter is quickly converted
by the hot-CNO cycle into mostly $^{14}$O and $^{15}$O. Using $^{14}$N
as a proxy for all accreted CNO has numerical advantages. Moreover,
all the CNO has to be assumed to have been processed to $^{14}$N
in the donor star prior to accretion in the case of enhanced $Y$
(e.g., Section \ref{sub:Models-with-composition}) or if the accretion
layer originates deep inside the donor. Keeping the same metal composition
for all models allows for easier comparison. For simplicity we do
not include other metals in the accretion composition.

\subsection{Models with composition variation\label{sub:Models-with-composition}}

To study the effect of the accretion composition on $\dot{M}_{\mathrm{st}}$,
we employ a set of simulations that was created with an earlier version
of KEPLER. Some of the simulations were presented in previous studies
\citep{Woosley2004,Heger2005,Heger2007}. The basic setup of the models
is the same as for the previously discussed set (Section \ref{sub:Models-with-rate}),
with the exception of the following. 

Thermonuclear rates are used from a compilation by \citet{Rauscher2003}.
In particular, $^{15}\mathrm{O}\left(\alpha,\gamma\right)\mathrm{^{19}Ne}$
and $3\alpha$ are taken from CF88, and $^{18}\mathrm{Ne}\left(\alpha,p\right)\mathrm{^{21}Na}$
from \citeauthor{Rauscher2000} (2000, RATH; Figures~\ref{fig:rate_o15ag}
and \ref{fig:rate_ne18ap}). 

For these models the outer zone has a mass of $2\times10^{19}\,\mathrm{g}$,
which is well above the depth where H/He burning takes place, and
$Q_{\mathrm{b}}=0.15\,\mathrm{MeV\, nucleon^{-1}}$. Accretion is
simulated by increasing the pressure at the outer boundary each time
step by the weight of the newly accreted material until enough mass
for a new zone has accumulated. Then an extra zone is added at the
outside of the grid. The addition of a zone induces a brief dip in
the light curve as the zone is added with the same temperature as
the previous outermost zone and the temperature structure of the model
has to adjust. We carefully check that these dips do not influence
our results.

We create models for several values of the metallicity of the accreted
material, $Z$ ($^{14}$N), to represent the initial stellar abundances
of donor stars of different metallicity. We determine the $^{4}$He
mass fraction from the crude scaling relation $Y=0.24+1.76Z$, such
that for $Z=0$ the Big Bang Nucleosynthesis value is obtained, and
for $Z=0.02$ the solar value is reproduced (Section~\ref{sub:Models-with-rate};
see, e.g., \citealt{West2013} and references therein for more advanced
composition scaling relations). The remainder of the composition is
$^{1}$H: $X=1-Y-Z$. One may rewrite these equations to obtain a
relation between $X$ and $Z$: 
\begin{equation}
Z=0.362\left(0.76-X\right).\label{eq:X-Z}
\end{equation}
Our metallicity range extends up to $10$ times solar, which may be
applicable to some extreme cases towards the Galactic Bulge, or cases
of binary mass transfer from the evolved primary star (now the neutron
star) to the companion (now the donor). In Section \ref{sub:Accretion-Composition}
we compare the results of the two sets of models and confirm their
consistency.

\section{Results}

\subsection{Nuclear reaction rate dependence\label{sub:Nuclear-reaction-rate}}

We perform $243$ simulations to study the dependence of $\dot{M}_{\mathrm{st}}$
on three key reaction rates. We take steps in $\dot{M}$ to locate
the transition between stable and unstable burning. The smallest steps
are $0.01\,\dot{M}_{\mathrm{Edd}}$. Figure~\ref{fig:standard_lcv}
shows for the standard reaction rates two light curves of models around
the stability transition that differ in $\dot{M}$ by the smallest
step: $\dot{M}=1.07\,\dot{M}_{\mathrm{Edd}}$ (top panel) and $\dot{M}=1.08\,\dot{M}_{\mathrm{Edd}}$
(bottom panel). The light curves include the start of the simulations,
when the nuclear burning has not settled in its final behavior. The
first burst is more energetic, as the subsequent bursts ignite in
an environment rich in burst ashes \citep[compositional inertia,][]{Taam1980}.
The bursts appear to have an extended tail, as some of the freshly
accreted fuel prolongs the burning: approximately $75\%$ of the fluence
is emitted in the long tail, where we define the start of the tail
where the flux drops below $25\%$ of the value at the burst peak.
For the model with $\dot{M}=1.07\,\dot{M}_{\mathrm{Edd}}$, the $\alpha$-parameter
measures on average $\alpha=96$.

\subsubsection{Convergence of burning behavior}

The transition takes place in a small interval of $\dot{M}$. Because
we initialize the models without nuclear burning, the first simulated
bursts heat up the model slightly, which changes the burning burning
behavior somewhat. When we are close to the transition, this initial
heating changes the behavior to stable burning. For $\dot{M}=1.09\,\dot{M}_{\mathrm{Edd}}$
$2$ bursts appear before burning becomes stable, whereas for $\dot{M}=1.08\,\dot{M}_{\mathrm{Edd}}$
this number increases to $6$. We continue the simulation with $\dot{M}=1.07\,\dot{M}_{\mathrm{Edd}}$
until $40$ flashes are produced. There is variation in the recurrence
time of subsequent bursts: the fractional burst-to-burst variation
drops below $5\,\%$ after $5$ bursts, and afterwards the mean variation
is $1.4\,\%$. In principle the burning might still turn stable after
a larger number of bursts. The steepness of the increase with $\dot{M}$
of the number of bursts before stable burning, however, and the drop
in recurrence time variations after $5$ bursts, give us confidence
that we are close to the true transition when we choose to create
at least $12$ flashes per simulation. Simulations with stable burning
are continued for a similar physical time as simulations with bursts.

\begin{figure}
\includegraphics{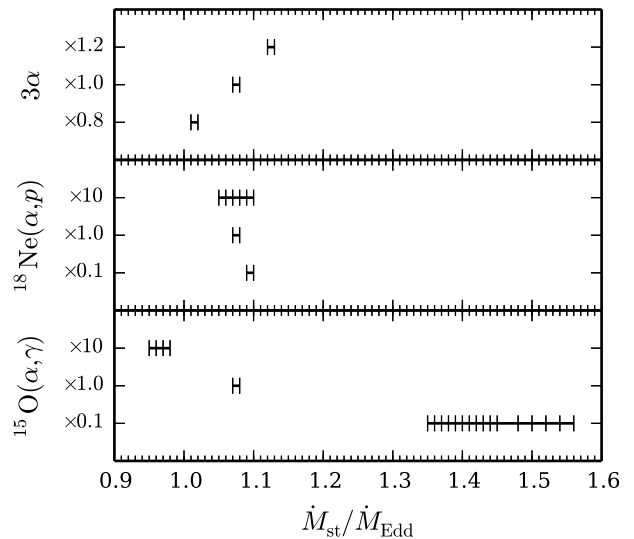}

\caption{\label{fig:overview_rates}For different reaction rate variations,
the mass accretion rates where models transition from bursts to stable
burning, $\dot{M}_{\mathrm{st}}$. Left side of each range is the
last bursting model, and the right side is the first stable model.
Tick marks in between indicate models with marginally stable burning.}
\end{figure}

\subsubsection{Variations with reaction rates}

\begin{figure}
\includegraphics{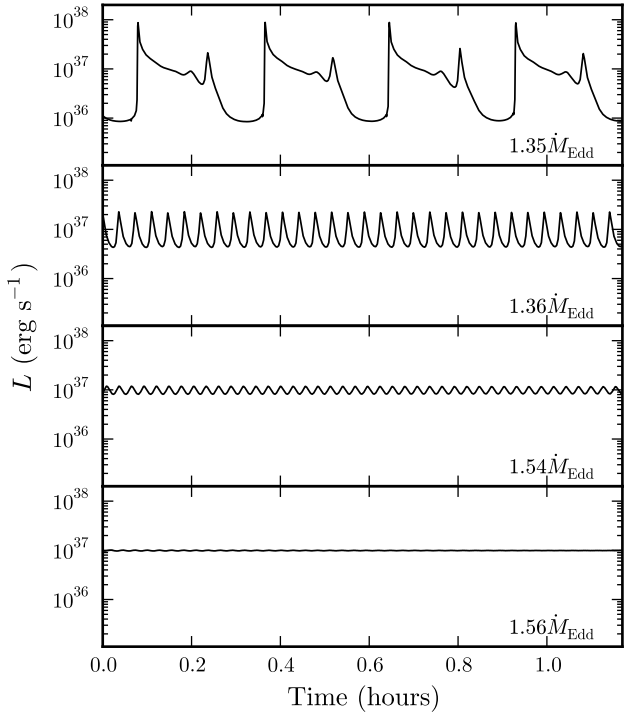}

\caption{\label{fig:o15ag_down_lcv}Light curves for models close to and at
the stability transition, with the decreased $^{15}\mathrm{O}\left(\alpha,\gamma\right)\mathrm{^{19}Ne}$
rate and mass accretion rates indicated as a multiple of $\dot{M}_{\mathrm{Edd}}$.
The full simulations last $6.9$ hours each.}
\end{figure}

The change in $\dot{M}_{\mathrm{st}}$ is at most $10\%$, with the
exception of the simulations where the $^{15}\mathrm{O}\left(\alpha,\gamma\right)\mathrm{^{19}Ne}$
rate is scaled by $0.1$: $\dot{M}_{\mathrm{st}}$ is increased by
$35\,\%$ (Figure~\ref{fig:overview_rates}). The interval $\Delta\dot{M}_{\mathrm{st}}$
where burning is marginally stable increases for both variations of
the $^{15}\mathrm{O}\left(\alpha,\gamma\right)\mathrm{^{19}Ne}$ rate
and for the increased $^{18}\mathrm{Ne}\left(\alpha,p\right)\mathrm{^{21}Na}$
rate. In the other series of simulations $\Delta\dot{M}_{\mathrm{st}}$
is smaller than our minimum step size. The $^{15}\mathrm{O}\left(\alpha,\gamma\right)\mathrm{^{19}Ne}$
downward variation yields the largest $\Delta\dot{M}_{\mathrm{st}}$,
which allows us to study in detail the changes in the burning behavior
and the corresponding light curves (Figure~\ref{fig:o15ag_down_lcv}).
Similar to Figure~\ref{fig:standard_lcv} the burning behavior changes
from bursts with extended tails to stable burning. In between, at
the boundary of stability, marginally stable burning produces oscillations
in the light curve. In this regime the onset of runaway burning is
repeatedly quenched when the cooling rate catches up with the burning
rate. At higher $\dot{M}$ the oscillations have a smaller amplitude
and are more symmetric. 

\begin{figure}
\includegraphics{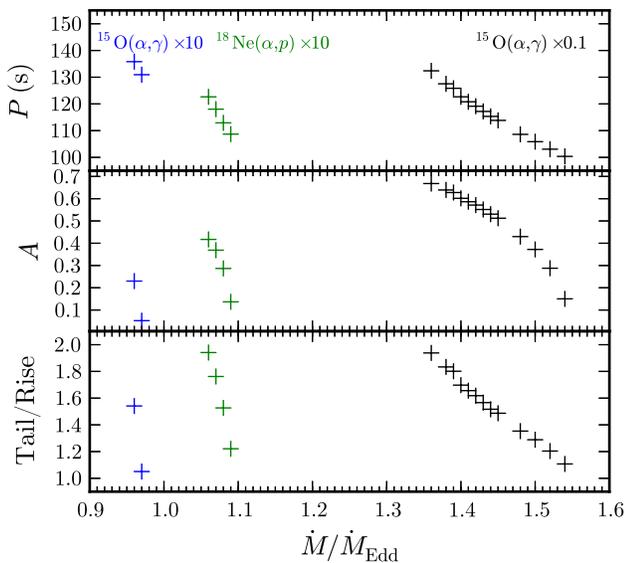}

\caption{\label{fig:osc_analyses}For the three series of models that exhibit
oscillations (Figure \ref{fig:overview_rates}) the period $P$, relative
amplitude $A$, and the ratio of the tail and rise times as a function
of mass accretion rate $\dot{M}$. Crosses only indicate the location
of data points; no uncertainty is implied.}
\end{figure}

We quantify the properties of the oscillations in the light curves
of all simulations with marginally stable burning by determining the
period, $P$, the relative amplitude, $A$, and the ratio of the duration
of the tail and the rise (Figure~\ref{fig:osc_analyses}). $A$ is
defined as half the luminosity difference between the maxima and minima,
normalized by their mean ($A=1$ when the oscillations account for
$100\,\%$ of the luminosity). The duration of the rise is the time
from a luminosity minimum to the next maximum; the tail duration is
analogously defined. At higher $\dot{M}$ both $P$ and $A$ are lower,
and the oscillations become more symmetric. The trends are similar
for all three series of simulations where we find marginally stable
burning.

\subsubsection{Nuclear burning at the stability transition\label{sub:Nuclear-burning-at}}

\begin{figure*}
\begin{centering}
\includegraphics{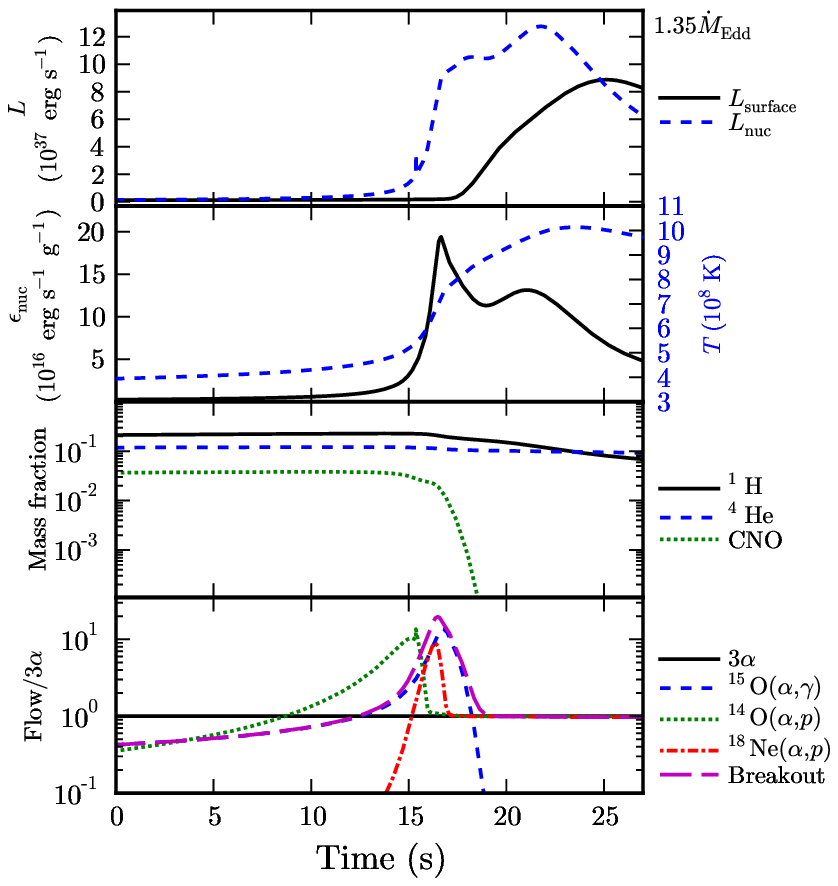} \includegraphics{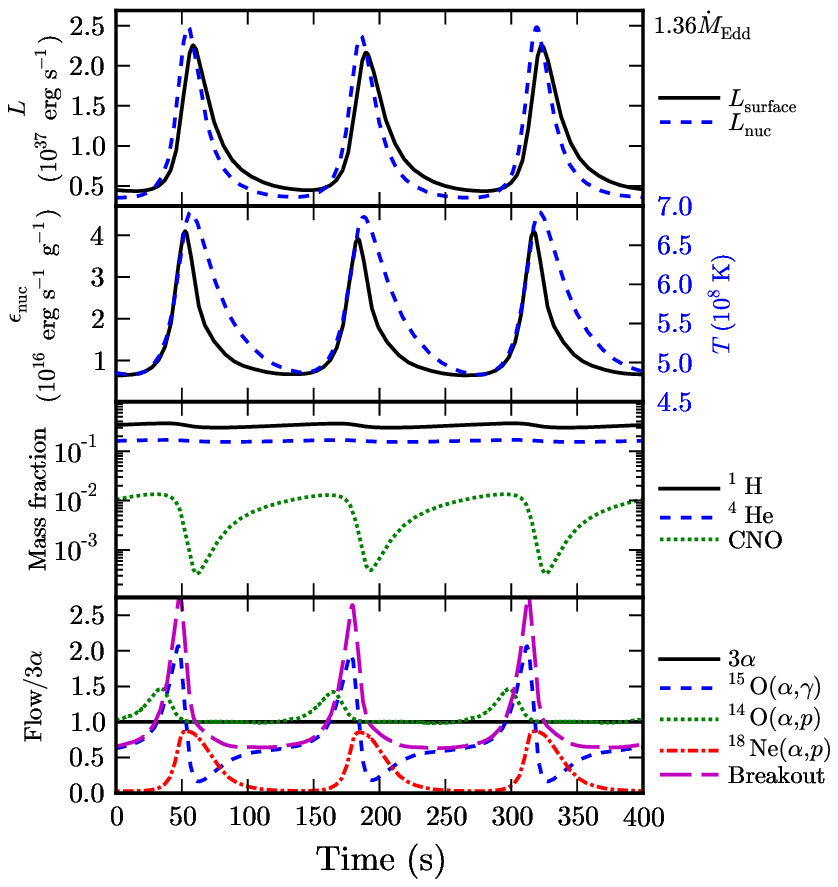} \includegraphics{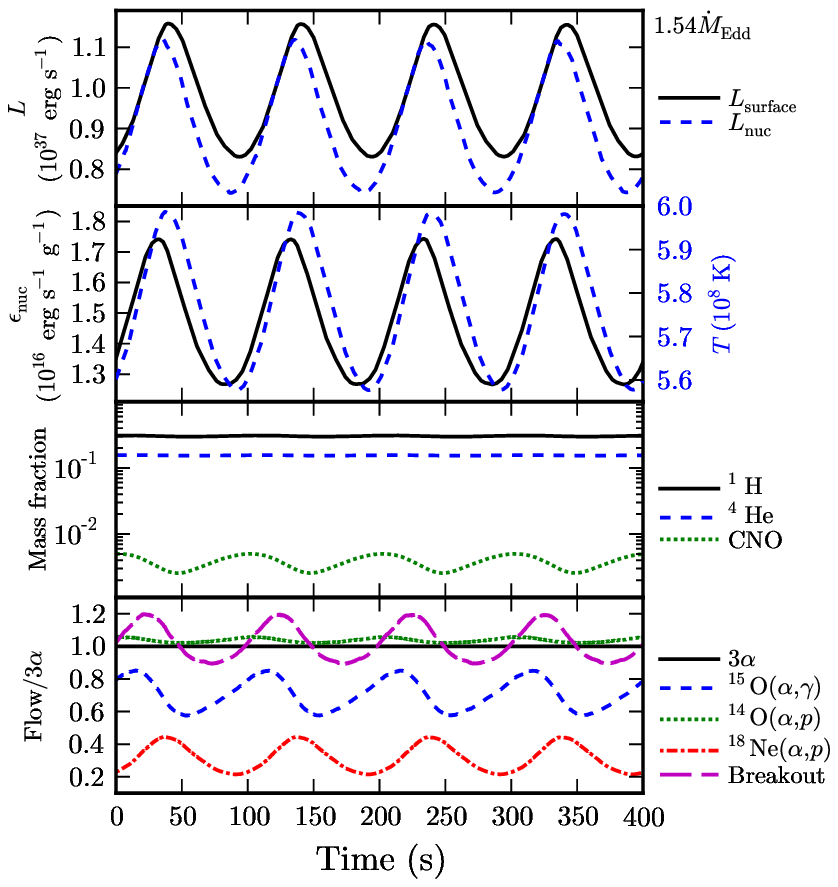}
\includegraphics{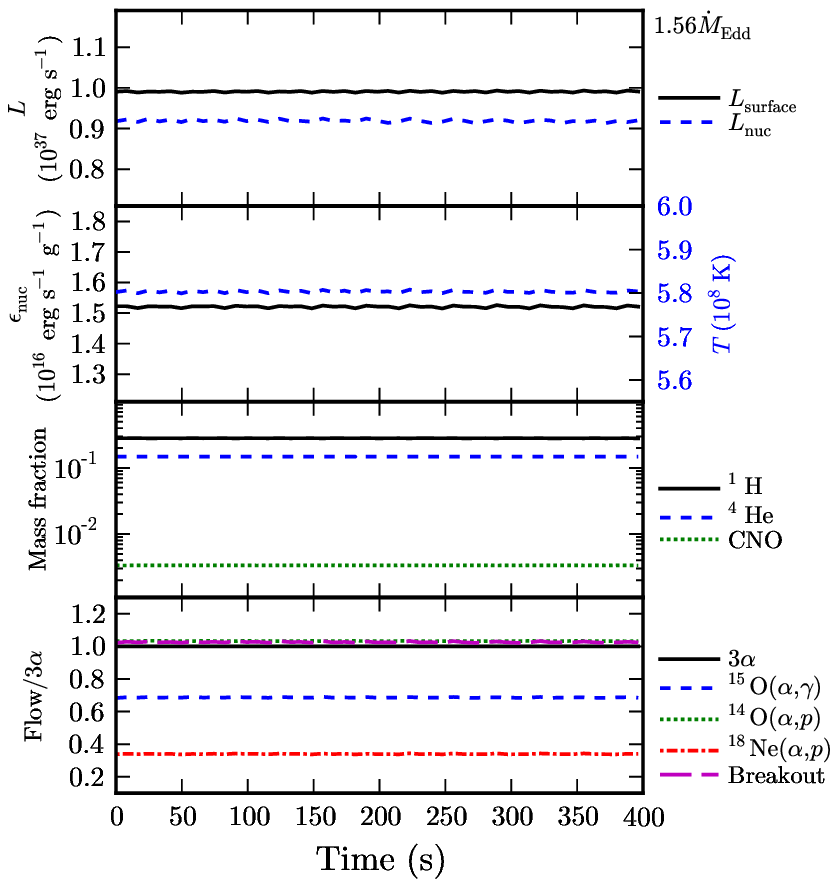}
\par\end{centering}

\caption{\label{fig:osc_flow_26}From the series of simulations with the decreased
$^{15}\mathrm{O}\left(\alpha,\gamma\right)\mathrm{^{19}Ne}$ rate
we show details for all models in Figure~\ref{fig:o15ag_down_lcv},
with $\dot{M}$ indicated in the top right of each panel. For each
model, from top to bottom, we give the surface luminosity $L$ and
total nuclear energy generation rate $L_{\mathrm{nuc}}$, as well
as at the depth where the time-averaged specific nuclear energy generation
rate $\epsilon_{\mathrm{nuc}}$ is maximal, $\epsilon_{\mathrm{nuc}}$,
temperature $T$ (dashed line), hydrogen, helium, and CNO mass fractions,
and several important nuclear flows relative to the $3\alpha$ flow,
where `breakout' is the sum of $^{15}\mathrm{O}\left(\alpha,\gamma\right)\mathrm{^{19}Ne}$
and $^{18}\mathrm{Ne}\left(\alpha,p\right)\mathrm{^{21}Na}$.}
\end{figure*}
 We study in detail four models from the series with the reduced $^{15}\mathrm{O}\left(\alpha,\gamma\right)\mathrm{^{19}Ne}$
rate around the stability transition (Figures~\ref{fig:o15ag_down_lcv}
and \ref{fig:osc_flow_26}). The highest accretion rate at which we
find bursts is $\dot{M}=1.35\,\dot{M}_{\mathrm{Edd}}$, and $\dot{M}=1.56\,\dot{M}_{\mathrm{Edd}}$
is the lowest rate with stable burning. Burning is marginally stable
between $\dot{M}=1.36\,\dot{M}_{\mathrm{Edd}}$ and $\dot{M}=1.54\,\dot{M}_{\mathrm{Edd}}$
(Figure~\ref{fig:overview_rates}). Apart from the surface luminosity,
$L$, and the total nuclear energy generation rate (neutrino losses
subtracted), $L_{\mathrm{nuc}}$, we study several quantities at a
fixed column depth where the time-averaged $\epsilon_{\mathrm{nuc}}$
(the specific nuclear energy generation rate; neutrino losses subtracted)
is maximal, which is close to the bottom of the hydrogen layer. For
the different models  this location is between column depths $y=6.2\times10^{7}\,\mathrm{g\, cm^{-2}}$
and $y=1.1\times10^{8}\,\mathrm{g\, cm^{-2}}$. 

The four simulations have several features in common: there is a delay
of several seconds between $L_{\mathrm{nuc}}$ and $L$, because the
surface responds on a thermal timescale to changes in the nuclear
burning at the bottom of the fuel layer. $L_{\mathrm{nuc}}$ is typically
lower than $L$ by $\sim7\,\%$. The extra emitted energy comes from
compressional heating and crustal heating. These contributions are
substantial because of the high mass accretion rates. There is also
a short delay between $\epsilon_{\mathrm{nuc}}$ and $L_{\mathrm{nuc}}$
because the latter includes nuclear energy generation in neighboring
zones to which the burning spreads. Burning in neighboring zones also
causes $T$ to remain high for several seconds when $\epsilon_{\mathrm{nuc}}$
starts to decline. 

\begin{figure}
\includegraphics{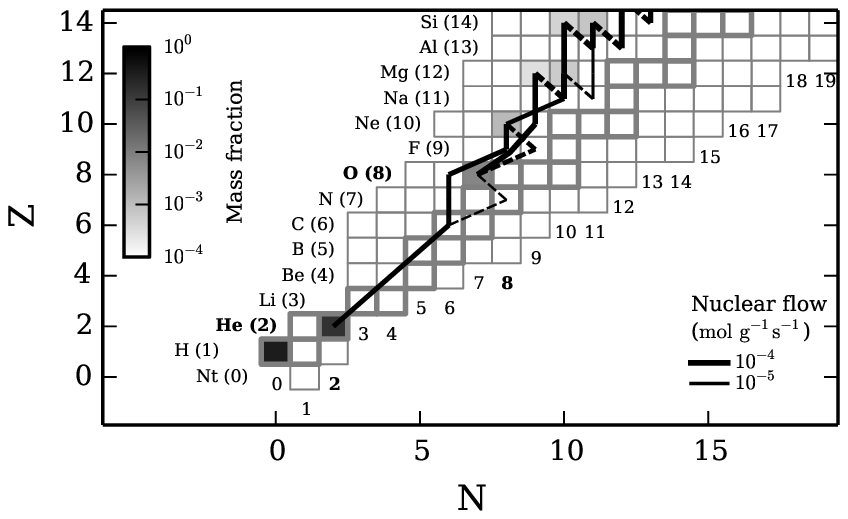}

\caption{\label{fig:flow_diagram}Net nuclear reaction flow for part of our
network during marginally stable burning (at $t=177\,\mathrm{s}$
for $\dot{M}=1.36\,\dot{M}_{\mathrm{Edd}}$ in Figure~\ref{fig:osc_flow_26}
at the depth where the time-averaged specific nuclear energy generation
is maximal). Each square is an isotope with neutron number $N$ and
charge number $Z$, with a color indicative of its mass fraction;
stable isotopes have thick borders. Lines indicate the flow by nuclear
reactions: solid lines flow to higher $Z$ and dashed lines to lower
$Z$. A continuous logarithmic distribution of line widths represents
the strength of the flows down to $1\%$ of the maximum; the legend
shows two line widths as example.}
\end{figure}

Next we discuss the burning behavior and reaction flows of each simulation.
Figure~\ref{fig:flow_diagram} shows the nuclear flow from the model
with $\dot{M}=1.36\,\dot{M}_{\mathrm{Edd}}$, but the path of the
flow is instructive for understanding all four models. The diagram
shows $^{4}\mathrm{He}$ burning to $^{12}\mathrm{C}$ through the
$3\alpha$ process; $^{1}\mathrm{H}$ burns via the hot CNO cycle,
which is extended with the bicycle through $^{14}\mathrm{O}\left(\alpha,p\right)^{17}\mathrm{F}$;
breakout from this cycle occurs via $^{15}\mathrm{O}\left(\alpha,\gamma\right)\mathrm{^{19}Ne}$
and $^{18}\mathrm{Ne}\left(\alpha,p\right)\mathrm{^{21}Na}$. From
$^{21}\mathrm{Na}$ there is no net flow back to the CNO cycle, and
the nuclear reactions continue towards heavier isotopes through the
\textsl{rp}-process (proton captures and $\beta^{+}$-decays). The
nuclear flow of Type I bursts has been studied in great detail before
\citep[e.g.,][]{Woosley2004,Fisker2008,Jose2010}. Here we investigate
the part that is responsible for the stability of the burning processes.

For $\dot{M}=1.35\,\dot{M}_{\mathrm{Edd}}$ burning is unstable. Here
we only study the burst onset when the thermonuclear runaway and the
breakout from the CNO cycle ensues (see \citealt{Woosley2004} for
a comprehensive study of X-ray burst models with the KEPLER code).
The burst starts with thermonuclear runaway burning of helium in the
$3\alpha$-process, raising the temperature such that an increasing
part of the nuclear flow proceeds through $^{14}\mathrm{O}\left(\alpha,p\right)^{17}\mathrm{F}$.
Note that we show the flows relative to the $3\alpha$ flow (Figure~\ref{fig:osc_flow_26}).
When $T>4.5\times10^{8}\,\mathrm{K}$, the breakout flow from the
hot CNO cycle via $^{15}\mathrm{O}\left(\alpha,\gamma\right)\mathrm{^{19}Ne}$
exceeds the $3\alpha$ flow: CNO is destroyed faster than it is created,
and its mass fraction starts to decline. The temperature continues
to increase, and when $T\simeq6.2\times10^{8}\,\mathrm{K}$ the breakout
flow through $^{18}\mathrm{Ne}\left(\alpha,p\right)\mathrm{^{21}Na}$
equals that of $^{15}\mathrm{O}\left(\alpha,\gamma\right)\mathrm{^{19}Ne}$.
Within seconds CNO is depleted. Nuclear burning continues with the
\textsl{$\alpha$p}- and \textsl{rp}-processes. The \textsl{rp}-process
waiting point at $^{30}$S produces a dip in $\epsilon_{\mathrm{nuc}}$,
which is visible as a `shoulder' in $L$ \citep[see also][]{Woosley2004}.

For stable burning at $\dot{M}=1.56\,\dot{M}_{\mathrm{Edd}}$ the
temperature in the burning layer is $T=5.8\times10^{8}\,\mathrm{K}$.
At this temperature both $^{15}\mathrm{O}\left(\alpha,\gamma\right)\mathrm{^{19}Ne}$
and $^{18}\mathrm{Ne}\left(\alpha,p\right)\mathrm{^{21}Na}$ contribute
substantially to the CNO breakout, with the former being the largest.
The breakout flow equals the $3\alpha$ flow: CNO is destroyed as
quickly as it is produced. In the models of marginally stable burning
($\dot{M}=1.36\,\dot{M}_{\mathrm{Edd}}$ and $\dot{M}=1.54\,\dot{M}_{\mathrm{Edd}}$)
the various quantities oscillate around the corresponding values in
the stable burning model. 

For $\dot{M}=1.36\,\dot{M}_{\mathrm{Edd}}$ the oscillations in the
$3\alpha$ nuclear flow follow the changes in $T$. In the minima
the $^{14}\mathrm{O}\left(\alpha,p\right)^{17}\mathrm{F}$ flow follows
$3\alpha$ closely. The $3\alpha$ reactions enlarge the CNO mass
fraction, which increases the $^{14}\mathrm{O}\left(\alpha,p\right)^{17}\mathrm{F}$
flow. This means hydrogen is burned at an increasing rate in the hot
CNO cycle as long as the flow through the breakout reactions is relatively
small. Once $T$ increases, the $^{15}\mathrm{O}\left(\alpha,\gamma\right)\mathrm{^{19}Ne}$
and $^{18}\mathrm{Ne}\left(\alpha,p\right)\mathrm{^{21}Na}$ flows
increase: the former becomes larger than $3\alpha$ and the CNO mass
fraction declines. This strongly reduces the $^{15}\mathrm{O}\left(\alpha,\gamma\right)\mathrm{^{19}Ne}$
flow and returns $^{14}\mathrm{O}\left(\alpha,p\right)^{17}\mathrm{F}$
to trace $3\alpha$. When the $^{18}\mathrm{Ne}\left(\alpha,p\right)\mathrm{^{21}Na}$
flow exceeds the $^{15}\mathrm{O}\left(\alpha,\gamma\right)\mathrm{^{19}Ne}$
flow, $\epsilon_{\mathrm{nuc}}$ peaks and starts to decrease. This
is because the chain {\footnotesize{$^{18}\mathrm{Ne}(\beta^{+})^{18}\mathrm{F}(p,\alpha)^{15}\mathrm{O}(\alpha,\gamma)^{19}\mathrm{Ne}(p,\gamma)^{20}\mathrm{Na}(p,\gamma)^{21}\mathrm{Mg}(\beta^{+})^{21}\mathrm{Na}$}}
releases $20.6\,\mathrm{MeV}$, whereas the direct reaction $^{18}\mathrm{Ne}\left(\alpha,p\right)\mathrm{^{21}Na}$
generates a mere $2.6\,\mathrm{MeV}$. The difference is that the
former effectively converts protons into a $^{4}$He nucleus and releases
the mass difference between the protons and $^{4}$He, although some
energy is carried away by neutrinos and the two $\beta^{+}$-decays.
These decays also limit the speed at which the process can operate,
such that at higher $T$ it can no longer compete with the direct
reaction. Even though $T$ still increases for a brief time, $\epsilon_{\mathrm{nuc}}$
decreases. Once the combined flow through the breakout reactions is
reduced below $3\alpha$, the CNO mass fraction increases again. During
the oscillations the $^{1}\mathrm{H}$ fraction changes locally by
$26\,\%$ and the $^{4}\mathrm{He}$ fraction by $11\,\%$.

For $\dot{M}=1.54\,\dot{M}_{\mathrm{Edd}}$, during the oscillations
the $^{1}\mathrm{H}$ fraction changes locally by $3\,\%$ and the
$^{4}\mathrm{He}$ fraction by $2\,\%$. Similar to the model with
$\dot{M}=1.36\,\dot{M}_{\mathrm{Edd}}$, the CNO mass fraction grows
or shrinks depending on whether the $3\alpha$ or the breakout flow
is larger. Unlike that model, $T$ remains high enough all the time
such that $^{18}\mathrm{Ne}\left(\alpha,p\right)\mathrm{^{21}Na}$
is never switched off. The $^{18}\mathrm{Ne}\left(\alpha,p\right)\mathrm{^{21}Na}$
flow never exceeds $^{15}\mathrm{O}\left(\alpha,\gamma\right)\mathrm{^{19}Ne}$,
although the relative contribution of the two reactions to the total
breakout does change periodically. The result is oscillations in the
light curve that have a smaller amplitude and are more symmetric than
for $\dot{M}=1.36\,\dot{M}_{\mathrm{Edd}}$, which produces less symmetric
oscillations, because of the faster destruction of CNO by the breakout
reactions.

\subsubsection{Energy generation rate and composition}

\begin{figure}
\includegraphics{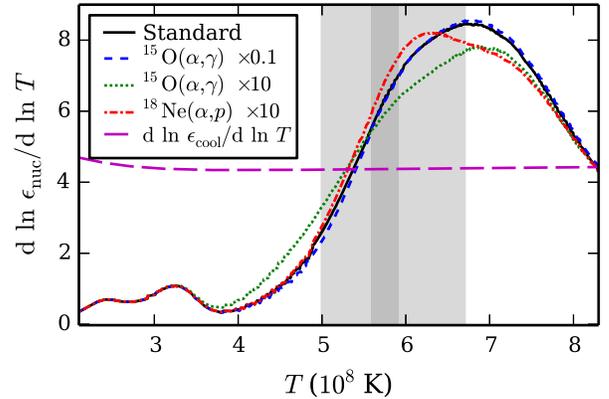}

\caption{\label{fig:eb}Temperature dependence of the specific nuclear energy
generation rate $\epsilon_{\mathrm{nuc}}$ for different sets of reaction
rates. The gray bands indicate the temperature ranges from two models
with marginally stable burning (Figure~\ref{fig:osc_flow_26}). }
\end{figure}
Stability of thermonuclear burning is often determined by comparing
the temperature dependence of the specific nuclear energy generation
rate, $\mathrm{d}\ln\epsilon_{\mathrm{nuc}}/\mathrm{d}\ln T$, to
that of the specific cooling rate, $\mathrm{d}\ln\epsilon_{\mathrm{cool}}/\mathrm{d}\ln T$
(e.g., \citealt{Bildsten1998}). From the simulations with the standard
rate set we select the stable burning model with the lowest $\dot{M}$
(Figure~\ref{fig:standard_lcv}). In the zone of maximal specific
energy generation, we calculate $\mathrm{d}\ln\epsilon_{\mathrm{nuc}}/\mathrm{d}\ln T$
for each set of reaction rates, such that in each calculation we use
the same composition and we only probe the changes because of the
reaction rates (Figure~\ref{fig:eb}). For the rate set with the
largest change in $\dot{M}_{\mathrm{st}}$, $^{15}\mathrm{O}\left(\alpha,\gamma\right)\mathrm{^{19}Ne}$
scaled by $0.1$, $\mathrm{d}\ln\epsilon_{\mathrm{nuc}}/\mathrm{d}\ln T$
is very close to the result for the standard rates, whereas larger
deviations are found for rate sets that have smaller changes in $\dot{M}_{\mathrm{st}}$.
We also calculate $\mathrm{d}\ln\epsilon_{\mathrm{cool}}/\mathrm{d}\ln T$,
which is close to $4.4$ in the temperature range of interest, slightly
higher than the value of $4.0$ expected from simple radiative cooling
with $\epsilon_{\mathrm{cool}}\propto T^{4}$ \citep[e.g.,][]{Bildsten1998}. 

\begin{figure}
\includegraphics{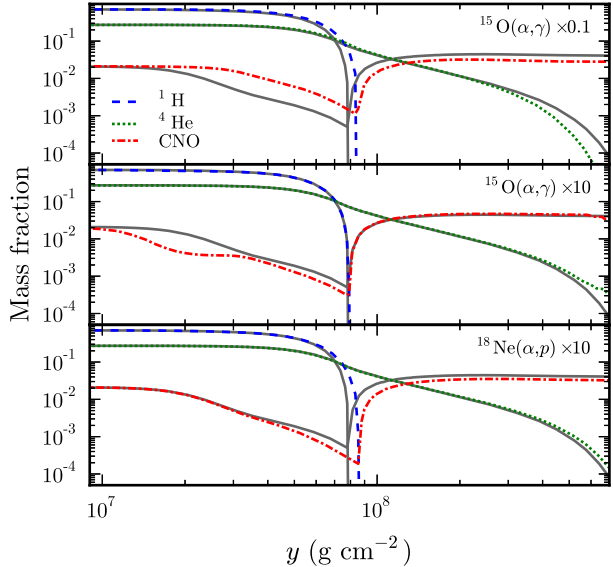}

\caption{\label{fig:composition_stable}Mass fractions of $^{1}$H, $^{4}$He,
and CNO as a function of column depth, $y$, for stable burning models
with different sets of reaction rates. Solid lines indicate the composition
for the standard set of reaction rates.}
\end{figure}
When reaction rates are changed, this has consequences for the composition
as a function of depth during stable burning. Figure~\ref{fig:composition_stable}
shows the differences in the composition profiles for the first models
with stable burning (Figure~\ref{fig:overview_rates}) for the reaction
rate variations that yield the largest differences in burning behavior
from our standard set of rates. 
\begin{figure}
\includegraphics{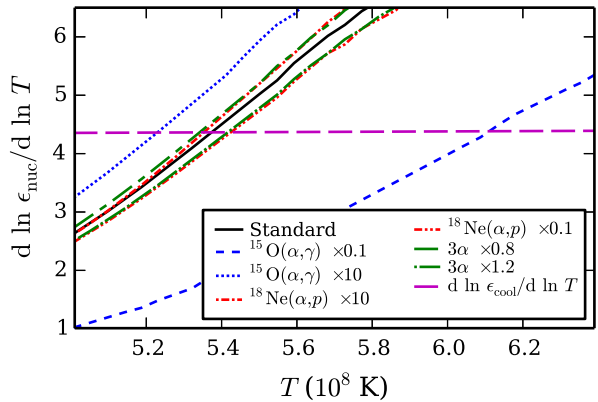}

\caption{\label{fig:eb_stable}Temperature dependence of the specific nuclear
energy generation rate $\epsilon_{\mathrm{nuc}}$ for different sets
of reaction rates zoomed-in on the stability transition. For each
rate variation we show the first model with stable burning (Figure~\ref{fig:overview_rates}),
which means each curve is for a different equilibrium composition
(Figure~\ref{fig:composition_stable}).}
\end{figure}
 For these and similar models for all other rate variations, we calculate
the temperature dependence of $\epsilon_{\mathrm{nuc}}$ (Figure~\ref{fig:eb_stable}).
The cooling rate's temperature sensitivity, $\mathrm{d}\ln\epsilon_{\mathrm{cool}}/\mathrm{d}\ln T$,
is the same for all models within the temperature range of interest.
The range of values of $T$ where $\mathrm{d}\ln\epsilon_{\mathrm{nuc}}/\mathrm{d}\ln T=\mathrm{d}\ln\epsilon_{\mathrm{cool}}/\mathrm{d}\ln T$
is wider than when we calculated $\mathrm{d}\ln\epsilon_{\mathrm{nuc}}/\mathrm{d}\ln T$
with the same composition (Figure~\ref{fig:eb}). Different reaction
rates lead, therefore, to changes in the equilibrium composition for
stable burning, and the composition has a large influence on the stability
of nuclear burning. Note that we found for the reduced $^{15}\mathrm{O}\left(\alpha,\gamma\right)\mathrm{^{19}Ne}$
rate that the model with stable burning (Figure~\ref{fig:osc_flow_26},
$\dot{M}=1.56\,\dot{M}_{\mathrm{Edd}}$) has a temperature of $T=5.8\times10^{8}\,\mathrm{K}$
in the zone of maximal specific energy generation, whereas $\mathrm{d}\ln\epsilon_{\mathrm{nuc}}/\mathrm{d}\ln T=\mathrm{d}\ln\epsilon_{\mathrm{cool}}/\mathrm{d}\ln T$
at $T=6.1\times10^{8}\,\mathrm{K}$. This exemplifies the limitations
of using a one-zone criterion for determining the stability of thermonuclear
burning in a multi-zone model.

\subsection{Compositional dependence\label{sub:Compositional-dependence}}

\begin{figure}
\includegraphics{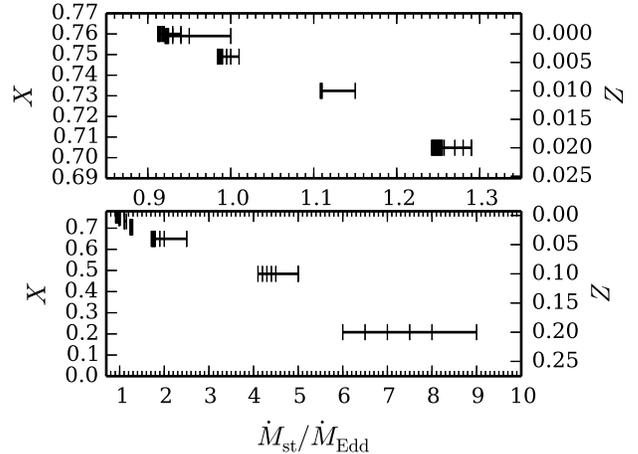}

\caption{\label{fig:overview_X}Similar to Figure~\ref{fig:overview_rates}
for different compositions indicated by $X$ and $Z$. The top panel
is a zoom-in of the bottom panel. In highly sampled regions the tick
marks are indistinguishable from each other.}
\end{figure}
\begin{figure}
\includegraphics[bb=183bp 310bp 428bp 480bp,clip]{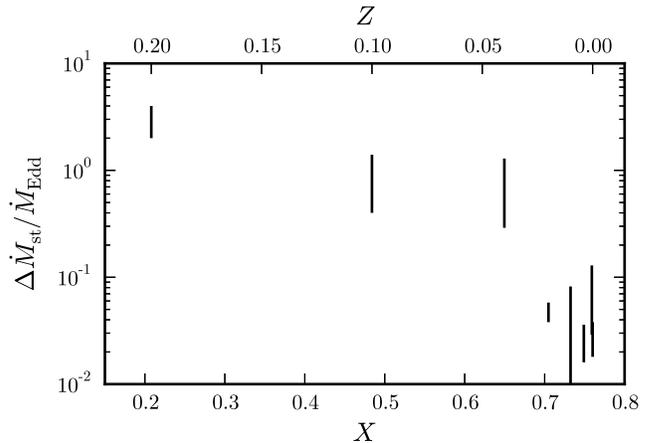}

\caption{\label{fig:width_X}For different compositions indicated by $X$ and
$Z$, the width of the mass accretion region where models transition
from bursts to stable burning, $\Delta\dot{M}_{\mathrm{st}}$. The
error bar reflects the maximum step size of the simulation grid (see
also Figure~\ref{fig:overview_X}).}
\end{figure}
The effect of varying the accretion composition is investigated using
a large set of $472$ simulations. The multi-zone simulations presented
by \citet{Heger2005} are included in this set ($X=0.759$, $Z=0.001$).
For combinations of decreasing $X$ and increasing $Z$ the transition
moves to higher values of $\dot{M}_{\mathrm{st}}$ (Figure~\ref{fig:overview_X}),
and the width of the transition, $\Delta\dot{M}_{\mathrm{st}}$, increases
(Figure~\ref{fig:width_X}). The trend in $\Delta\dot{M}_{\mathrm{st}}$
as a function of composition appears bimodal, as $\Delta\dot{M}_{\mathrm{st}}$
changes by an order of magnitude around $X\simeq0.65$.

\begin{figure}
\includegraphics{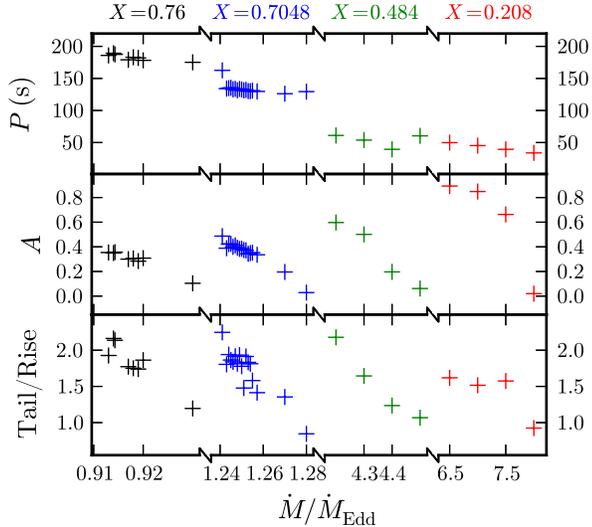}

\caption{\label{fig:osc_analyses_X}Same as Figure~\ref{fig:osc_analyses}
for four series of models with different accretion compositions as
indicated with the hydrogen mass fraction $X$. Note the breaks in
the horizontal axis.}
\end{figure}
 For four series of simulations (four accretion compositions) we study
the properties of the oscillations in the luminosity (Figure~\ref{fig:osc_analyses_X}).
These simulations suffer from small dips in the light curve, which
prevents us from determining the properties with the same precision
as for the simulations in Section~\ref{sub:Nuclear-reaction-rate}.
This produces a small amount of noise in Figure~\ref{fig:osc_analyses_X}.
Nevertheless, the trends in the properties are clearly similar to
those in the previous series of models (Figure~\ref{fig:osc_analyses}):
for larger $\dot{M}$, $P$ and $A$ are smaller, and the oscillations
are more symmetric. Values of $P$ range from $34\,\mathrm{s}$ to
$189\,\mathrm{s}$, and oscillations with vanishingly small amplitudes
are completely symmetric with respect to the rise and the tail. Furthermore,
models with a lower hydrogen mass-fraction produce oscillations with
on average a shorter period.

\subsection{Matching the observed transition}

None of the presented models reproduce the transition at the observed
$\dot{M}_{\mathrm{st}}$. We investigate whether extrapolation of
the trends implied by our models suggests that a certain reaction
rate or accretion composition does find $\dot{M}_{\mathrm{st}}=0.1\dot{M}_{\mathrm{Edd}}$.

We have performed simulations for three rate variations of each reaction
(Section~\ref{sub:Nuclear-reaction-rate}). Even though this is a
limited number, we use the implied trends to find how large a change
in the different rates is required to match observations. One needs
to either reduce the $3\alpha$ rate or increase the $^{18}\mathrm{Ne}\left(\alpha,p\right)\mathrm{^{21}Na}$
or $^{15}\mathrm{O}\left(\alpha,\gamma\right)\mathrm{^{19}Ne}$ rate
by over $4$ orders of magnitude (Figure~\ref{fig:overview_rates}).
This is far outside the values allowed by nuclear experiment.

A naive linear extrapolation of the models with different accretion
composition finds $\dot{M}_{\mathrm{st}}=0.1\dot{M}_{\mathrm{Edd}}$
is reproduced for $X\simeq0.9$. In our prescription for the composition
this implies a negative metallicity and, therefore, these solutions
are unphysical. 

In conclusion, although the extrapolations from the results of our
simulations are crude, they suggest that no allowed change in accretion
composition or reaction rate results in a stability transition at
the observed $\dot{M}_{\mathrm{st}}$.

\section{Discussion}

One of the largest challenges for the theory of thermonuclear burning
in the neutron star envelope is to resolve the discrepancy between
models/theory and observations on the mass accretion rate where stable
burning sets in, $\dot{M}_{\mathrm{st}}$. Using large sets of one-dimensional
multi-zone simulations we investigate the dependence of $\dot{M}_{\mathrm{st}}$
on the reaction rates of the $3\alpha$- and CNO breakout processes,
as well as on the accretion composition. Although we find that the
dependence can be strong, the simulations are unable to reproduce
the observed value of $\dot{M}_{\mathrm{st}}$.

\subsection{Reaction rates}

Hydrogen burns faster in the \textsl{rp}-process than in the $\beta$-limited
CNO cycle. A higher $3\alpha$ rate increases the $^{12}$C mass fraction
and boosts CNO cycle burning, reducing the fraction of $^{1}$H that
burns in the faster \textsl{rp}-process. Higher breakout rates, on
the other hand, reduce the CNO mass fraction, causing the opposite
effect. This results in an increased $\dot{M}_{\mathrm{st}}$ for
a larger $3\alpha$ rate, whereas the trend is reversed for the two
breakout reactions (Figure~\ref{fig:overview_rates}).

The only reaction rate variation that changes $\dot{M}_{\mathrm{st}}$
by more than $10\,\%$ is $^{15}\mathrm{O}\left(\alpha,\gamma\right)\mathrm{^{19}Ne}$
scaled by $0.1$, with $\dot{M}_{\mathrm{st}}=1.455\,\dot{M}_{\mathrm{Edd}}$.
This leads us further from the observed $\dot{M}_{\mathrm{st}}\simeq0.1\,\dot{M}_{\mathrm{Edd}}$.
The discrepancy with observations is, therefore, not resolved by the
uncertainty in the considered reaction rates, and it may be substantially
larger depending on the actual $^{15}\mathrm{O}\left(\alpha,\gamma\right)\mathrm{^{19}Ne}$
rate.

The important role of $^{15}\mathrm{O}\left(\alpha,\gamma\right)\mathrm{^{19}Ne}$
in the stability of thermonuclear burning has been highlighted in
several previous studies. \citet{Cooper2006} found from a stability
analysis that a reduced rate lowers $\dot{M}_{\mathrm{st}}$ substantially
\citep[see also][]{Cooper2006a}. This was, however, in the context
of so-called delayed-detonation bursts \citep{2003NarayanHeyl}, which
are not reproduced by multi-zone simulations (such as the ones in
this paper). 

In multi-zone models where the $^{15}\mathrm{O}\left(\alpha,\gamma\right)\mathrm{^{19}Ne}$
rate is orders of magnitude lower than what we considered (effectively
switching off CNO breakout), \citet{Fisker2006} identified a stable
burning regime over a wide range of $\dot{M}$ \citep[see also][]{Fisker2007}.
This precludes X-ray bursts from occurring at any $\dot{M}$. In an
attempt to reproduce this, \citet{Davids2011} performed a relatively
short multi-zone simulation with a reduced $^{15}\mathrm{O}\left(\alpha,\gamma\right)\mathrm{^{19}Ne}$
rate \citep[using a different implementation than][]{Fisker2006},
and found flashes instead of the stable burning. This regime is not
constrained by our models, because we focus on larger $\dot{M}$,
and we do not consider $^{15}\mathrm{O}\left(\alpha,\gamma\right)\mathrm{^{19}Ne}$
rates as low is in these studies.

Compared to \citet{Fisker2007}, we substantially improve on resolving
$\dot{M}_{\mathrm{st}}$ for three variations of the $^{15}\mathrm{O}\left(\alpha,\gamma\right)\mathrm{^{19}Ne}$
rate.

\subsection{Accretion Composition\label{sub:Accretion-Composition}}

Although the specific energy generation rate for hot CNO burning depends
solely on $Z$, the total nuclear energy generation rate increases
monotonically with $X$ for a given $\dot{M}$. Models with a lower
$X$ (larger $Y,\, Z$), therefore, need to make up for the lower
CNO-cycle heating rate with a larger $\dot{M}_{\mathrm{st}}$ (Figure~\ref{fig:overview_X}).
Extrapolation of the $\dot{M}_{\mathrm{st}}(X)$ trend suggests that
the observed value is reached for a metal-poor composition with $X\simeq0.9$.
Hydrogen mass fractions higher than the primordial value of $0.76$,
however, are not likely to occur in nature except if there is significant
spallation during the accretion process \citep{Bildsten1992}. Furthermore,
compositional inertia may preclude such a solution from working in
practice, as a burst ignites in the presence of the CNO-rich ashes
produced in the previous flash \citep{Taam1980,Woosley2004}.

\citet{Heger2005} employ one-zone calculations to study the effect
of both $X$ and the gravitational acceleration, $g$, on stability.
They argue that $Z$ has a weak effect on the transition accretion
rate, and, therefore, keep it at the solar value when varying $X$.
For two presented models with the same Newtonian gravitational acceleration
as our models ($g_{14}=1.9$), $\dot{M}_{\mathrm{st}}$ increases
by approximately a factor $1.8$ when changing from $X=0.7$ to $X=0.5$,
whereas a similar change in $X$ for our models increases the transitional
rate by a factor $3.6$ (Figure~\ref{fig:overview_X}). The one-zone
model with reduced $X$ has more helium than our corresponding multi-zone
model, so with respect to the energy that can be liberated by nuclear
burning of the accreted material, the difference between the models
with $X=0.7$ and $X=0.5$ is smaller for the one-zone than for our
multi-zone models. This may explain why the difference in transitional
mass accretion rate is also smaller for the one-zone models.

To check the self-consistency of the set of models with composition
variation and the set with reaction rate variation, consider the models
from the former with $X=0.7048$ and $Z=0.02$, which is close to
the composition used in the latter set. The two sets used different
prescriptions for the CNO breakout reaction rates (Figures~\ref{fig:rate_o15ag},
\ref{fig:rate_ne18ap}). Using simple interpolation of the $\dot{M}_{\mathrm{st}}$
values in Figure~\ref{fig:overview_rates} to derive a scaling for
$\dot{M}_{\mathrm{st}}$ of the composition variation set yields $\dot{M}_{\mathrm{st}}=1.23\,\dot{M}_{\mathrm{Edd}}$.
Keeping in mind the crudeness of this interpolation, this is reasonably
close to the value for the reaction rate variation set of $\dot{M}_{\mathrm{st}}=1.27\,\dot{M}_{\mathrm{Edd}}$.

\subsection{Width of the stability transition}

From the composition dependence, we find that for higher $Z$, $\Delta\dot{M}_{\mathrm{st}}$
increases, although the trend seems to change at $Z\simeq0.03$ (Figure~\ref{fig:width_X}),
indicating that the dependence is more complicated. For the rate variations,
however, the models with increased $\Delta\dot{M}_{\mathrm{st}}$
have both higher and lower $Z$ at the bottom of the hydrogen-rich
layer (Figure~\ref{fig:composition_stable}). Alternatively, we can
consider the temperature dependence of $\mathrm{d}\ln\epsilon_{\mathrm{nuc}}/\mathrm{d}\ln T$,
which for the models with a relatively wide $\Delta\dot{M}_{\mathrm{st}}$
is in two cases steeper and in one case shallower than for the standard
rate set. Therefore, $\Delta\dot{M}_{\mathrm{st}}$ is determined
by a more complex set of factors, which will require more detailed
study to unravel.

\citet{Heger2005}, using a one-zone model that only includes the
triple-$\alpha$ reaction, find $\Delta\dot{M}_{\mathrm{st}}\simeq0.01\,\dot{M}_{\mathrm{Edd}}$
for solar composition and the same gravitational acceleration as our
models. This is consistent with our models from the rate variation
study with the standard reaction rates.

When determining $\Delta\dot{M}_{\mathrm{st}}$ from X-ray observations,
the values of $\dot{M}$ that bound this interval may suffer from
substantial systematic uncertainties (Section~\ref{sec:Introduction}).
For a given source, however, both boundaries have the same systematic
error, and a meaningful value of $\Delta\dot{M}_{\mathrm{st}}$ can
be obtained nonetheless. The systematic uncertainty in $\Delta\dot{M}_{\mathrm{st}}$
is likely several tens of percents, the same as for $\dot{M}$ (Section~\ref{sec:Introduction}).
An additional problem is that when mHz QPOs are observed the accretion
rate may not be constant, and the burning may not have reached a limit
cycle, whereas our models represent equilibrium behavior at a constant
$\dot{M}$. For example, \citet{Altamirano2008} observed bursts and
mHz QPOs from 4U~1636--53 to alternate while the persistent flux
remained constant, and \citet{Keek2008} noted that bursts occurred
on 4U~1608--52 at accretion rates higher than those where mHz QPOs
are present. This makes the determination of $\Delta\dot{M}_{\mathrm{st}}$
from observations somewhat ambiguous. Nevertheless, based on observations
of mHz QPOs from 4U~1608--52 and 4U~1636--53, \citet{Revnivtsev2001}
find $\Delta\dot{M}_{\mathrm{st}}\simeq0.05\,\dot{M}_{\mathrm{Edd}}$,
which agrees with our prediction for solar composition accretion (Figure~\ref{fig:width_X}).
Compared to simulations with reaction rate variation, those models
had at the relevant temperatures a factor $\sim4$ times lower $^{18}\mathrm{Ne}\left(\alpha,p\right)\mathrm{^{21}Na}$
rate and $\sim2$ times lower $^{15}\mathrm{O}\left(\alpha,\gamma\right)\mathrm{^{19}Ne}$
rate (Figures~\ref{fig:rate_o15ag} and \ref{fig:rate_ne18ap}).
$\Delta\dot{M}_{\mathrm{st}}\simeq0.05\,\dot{M}_{\mathrm{Edd}}$ agrees
with the trend of larger $\Delta\dot{M}_{\mathrm{st}}$ for lower
$^{15}\mathrm{O}\left(\alpha,\gamma\right)\mathrm{^{19}Ne}$ rates
(Figure~\ref{fig:overview_rates}). 

For IGR~J17480--2446 \citet{Linares2011} identify mHz QPOs in a
range $\Delta\dot{M}_{\mathrm{st}}\simeq0.09\,\dot{M}_{\mathrm{Edd}}$.
In this case there is a smooth transition from bursts to QPOs, and
$\Delta\dot{M}_{\mathrm{st}}$ may have been over-estimated. Note
that the bursts from all mentioned X-ray sources indicate the accreted
material is hydrogen-rich.

\subsection{Marginally stable burning}

Analytic arguments, writing $\mathrm{d}\ln\epsilon_{\mathrm{nuc}}/\mathrm{d}\ln T=4+\beta$,
predict marginal stability when $\left|\beta\right|\lesssim10^{-2}$,
such that the `effective thermal timescale' is of similar size as
the accretion timescale \citep{Heger2005}. We find, however, that
during oscillatory burning the temperature and composition variations
produce values of $|\beta|$ of a few. The analytic arguments, therefore,
describe only very small perturbations from stability, whereas we
find oscillatory behavior to persist at larger perturbations. We find
that the marginally stable burning occurs because of a combination
of effects: the energy generation rate changes because of the destruction
and creation of CNO, as well as because of the changing path of the
nuclear flow through either of the hot-CNO breakout reactions. The
effective reduction of the energy generation as $T$ rises is, therefore,
larger than the increase in $\epsilon_{\mathrm{cool}}$ alone, which
may allow for larger $|\beta|$.

\citet{Keek2012} simulate hydrogen and helium burning in an atmosphere
that is cooling down from a superburst, and find a transition from
stable to marginally stable burning and bursts. The marginally stable
burning was found to be related to the switching on and off of the
$^{15}\mathrm{O}\left(\alpha,\gamma\right)\mathrm{^{19}Ne}$ breakout.
As in our simulations, the oscillatory burning is caused by the CNO
breakout reactions, but the details are different. The superburst
burst-quenching simulations produced oscillatory burning for a brief
time as the atmosphere cooled down, whereas in the current paper we
aim to model marginally stable burning for a longer time. The marginally
stable regime is approached differently in the two cases, which leads
to somewhat different behavior.

Over the range of considered accretion compositions, $\dot{M}_{\mathrm{st}}$
changes by a factor $8.1$ (Figure~\ref{fig:overview_X}). Because
of the importance of the accretion time scale on the period, $P$,
of marginally stable burning \citep{Heger2005}, this causes a wide
range of values for $P$ (Figure~\ref{fig:osc_analyses_X}). \citet{Altamirano2008}
observed several instances of mHz QPOs from 4U\,1636--53, where the
period of the oscillations increases over time until a Type I X-ray
burst occurred. In one case the period changed from $90\,\mathrm{s}$
to $130\,\mathrm{s}$. The width of this range is similar to the simulations
with the downward variation of the $^{15}\mathrm{O}\left(\alpha,\gamma\right)\mathrm{^{19}Ne}$
rate, although the values are somewhat lower when taking into account
a redshift of $z+1=1.26$, which can be explained by a smaller hydrogen
content.

\subsection{X-ray bursts with extended tails}

The bursts close to the transition have extended tails from the burning
of some freshly accreted fuel \citep{Heger2005}. The light curve
at the end of the tails may exhibit a few oscillations. The tails
extend for a substantial fraction of the burst recurrence time, and
during that phase up to $3$ times the fluence of the burst is emitted.
This is similar to a burst observed from GX~3+1, which exhibited
a $30\,\mathrm{min}$ extended tail after an initial $\lesssim10\,\mathrm{s}$
peak \citep{Chenevez2006}. The burst was observed when the accretion
rate was close to $0.1\, M_{\mathrm{Edd}}$, which is the observed
$\dot{M}_{\mathrm{st}}$.

If one were to include the emission in the extended tail as part of
the persistent emission, the $\alpha$-parameter would be several
times higher. Increases in $\alpha$ of this magnitude have been observed
close to the stability transition compared to bursts at lower $\dot{M}$,
and the value we obtain of $\alpha\simeq100$ is within the observed
range \citep{Paradijs1988,Cornelisse2003}.

\subsection{Alternative solutions}

We have demonstrated that uncertainties in neither the $3\alpha$
rate, the CNO break-out reaction rates, nor the accretion composition
can account for the discrepancy between the observed and predicted
value of $\dot{M}_{\mathrm{st}}$. Even the combination of the most
favorable composition and reaction rates is most likely insufficient.
Although we have not simulated such a configuration directly, the
changes towards lower $\dot{M}_{\mathrm{st}}$ produced by rate and
composition variations are orders of magnitude away from reaching
the observed $\dot{M}_{\mathrm{st}}$. 

Several alternative explanations have been put forward to reproduce
the observed value of $\dot{M}_{\mathrm{st}}$. \citet{Heger2005}
find $\dot{M}_{\mathrm{st}}$ to be proportional to the effective
gravity in the neutron star envelope, but a simple linear extrapolation
of those results suggests the observed value of $\dot{M}_{\mathrm{st}}$
cannot be obtained for physical values of the gravitational acceleration.
Another explanation is rotationally induced mixing or mixing due to
a rotationally induced magnetic field, where freshly accreted material
is quickly transported deeper where it can undergo steady-state burning
\citep{Piro2007,Keek2009}. If the mixing is too strong, however,
burst recurrence times of minutes are predicted \citep{Piro2007},
which have only been observed from the atypical burster IGR~J17480--2446
\citep{Linares2011}.

It has been suggested that the theoretical value of $\dot{M}_{\mathrm{st}}$
represents a local value at one spot on the neutron star surface \citep{Heger2005}.
This may be the case if accreted matter is funneled to the magnetic
poles. With the exception of the accretion-powered X-ray pulsars,
however, the magnetic field in most accreting LMXBs is thought to
be weak. A weak field is unable to confine the accreted fuel at the
poles down to the burst ignition depth \citep{Bildsten1997}, and
the fuel spreads across the surface on timescales much shorter than
the burst recurrence.

The most promising solution is an increased heat flux into the atmosphere
\citep{Keek2009}, possibly generated by pycnonuclear and electron
capture reactions in the crust \citep[e.g.,][]{Haensel2003,Gupta2007}
or by the dissipation of rotational energy through turbulent braking
at the envelope-crust interface \citep{Inogamov2010}. This heat flux
is tempered by neutrino cooling in the outer crust \citep{Schatz2013},
and both heating and cooling sources will need to be carefully balanced
to reconcile simulations with the observed $\dot{M}_{\mathrm{st}}$.

\section{Conclusions}

Using large series of one-dimensional multi-zone simulations, we investigate
the dependence of the transition of stability of thermonuclear burning
on neutron stars on the reaction rates of the triple-alpha reaction
and the hot-CNO cycle breakout reactions $^{15}\mathrm{O}\left(\alpha,\gamma\right)\mathrm{^{19}Ne}$
and $^{18}\mathrm{Ne}\left(\alpha,p\right)\mathrm{^{21}Na}$. Within
the nuclear experimental uncertainties of the rates, a reduction of
the $^{15}\mathrm{O}\left(\alpha,\gamma\right)\mathrm{^{19}Ne}$ by
a factor $0.1$ produces the largest change in the mass accretion
rate where stability changes: $\dot{M}_{\mathrm{st}}$ is increased
from $1.08\,\dot{M}_{\mathrm{Edd}}$ to $1.46\,\dot{M}_{\mathrm{Edd}}$.
The lowest value of $\dot{M}_{\mathrm{st}}=0.97\,\dot{M}_{\mathrm{Edd}}$
is obtained for an increased $^{15}\mathrm{O}\left(\alpha,\gamma\right)\mathrm{^{19}Ne}$
rate by a factor $10$. Within the current nuclear uncertainties we
are, therefore, unable to explain the discrepancy with observations,
which find $\dot{M}_{\mathrm{st}}\simeq0.1\,\dot{M}_{\mathrm{Edd}}$.

We also study the dependence of $\dot{M}_{\mathrm{st}}$ on the accretion
composition. Reducing the hydrogen mass fraction below the solar value
increases $\dot{M}_{\mathrm{st}}$, leading it further away from the
observed value. An additional effect is the increase of the accretion
rate interval $\Delta\dot{M}_{\mathrm{st}}$ where burning is marginally
stable. For several reaction rate variations $\Delta\dot{M}_{\mathrm{st}}$
increases as well. $\Delta\dot{M}_{\mathrm{st}}$ appears to have
a complex dependence on the different reaction rates and the composition,
which requires further study to determine.

Close to the stability transition, we identify X-ray bursts with extended
tails lasting over $10$ minutes, where freshly accreted material
continues the nuclear burning.

Our simulations yield values of $\Delta\dot{M}_{\mathrm{st}}$, of
the marginally stable burning period, and of the $\alpha$-parameter
that are consistent with observations. Because of the dependency of
these parameters on $\dot{M}_{\mathrm{st}}$, however, quantitative
comparisons are problematic as long as the observed $\dot{M}_{\mathrm{st}}$
is not reproduced. Furthermore, given the degeneracy in many of these
parameters with respect to variations in reaction rates, accretion
composition, as well as the effective surface gravity, it remains
challenging to place constraints with current observations.

\acknowledgements{The authors thank the International Space Science Institute in Bern
for hosting an International Team on Type I X-ray bursts. LK and RHC
are supported by the Joint Institute for Nuclear Astrophysics (JINA;
grant PHY08-22648), a National Science Foundation Physics Frontier
Center. AH is supported by an ARC Future Fellowship (FT120100363).}

\bibliographystyle{apj}
\bibliography{apj-jour,stability}

\end{document}